# This Class Isn't Designed For Me: Recognizing Ableist Trends In Design Education, And Redesigning For An Inclusive And Sustainable Future

Sourojit Ghosh[a], Sarah Coppola[a]

[a]Human Centered Design and Engineering, University of Washington, Seattle

*Corresponding author e-mail: ghosh100@uw.edu

**Abstract**: Traditional and currently-prevalent pedagogies of design perpetuate ableist and exclusionary notions of what it means to be a designer. In this paper, we trace such historically exclusionary norms of design education, and highlight modern-day instances from our own experiences as design educators in such epistemologies. Towards imagining a more inclusive and sustainable future of design education, we present three case studies from our own experience as design educators in redesigning course experiences for blind and low-vision (BLV), deaf and hard-of-hearing (DHH) students, and students with other disabilities. In documenting successful and unsuccessful practices, we imagine what a pedagogy of care in design education would look like.

## 1. Introduction

Design, and design thinking has proliferated through multiple disciplines, including business and engineering, far beyond its original roots as a creative practice (Dreyfuss, 1955; Vinsel, 2020). In recent years, there has been serious critique of the ways in which the designed world excludes and harms people based on identity factors such as race, gender, nationality, and disability (e.g. Benjamin, 2019; Broussard, 2023; Holmes, 2020; Noble, 2018). This has led to a call to train designers who think and work more inclusively, and for initiatives that diversify the design workforce (Spiel & Angelini, 2022). However, the traditional pedagogies of design, such as long face-to-face studios or co-design sessions strongly focused on visual attributes of design, continue to exclude students and reify who gets to be and is considered a "designer". In this paper, we will discuss case studies from interdisciplinary design classes, as we highlight the specific ways in which existing design pedagogies are rife with ableist and exclusionary practices that affect students with traditionally marginalized identities. We document our own practices in redesigning course components to make our classrooms more inclusive, discuss successes and failures, and consider how such practices could ultimately contribute towards establishing a pedagogy of care in design education.

## 2. Ableist and Exclusionary Practices in Design Education

In recent years, there has been interest in examining the impact of design (Chardin & Novak, 2020) as well as a call for a commitment to design that advances diversity, equity, and inclusion (Costanza-Chock, 2020; Noel, 2016; Roscoe et al., 2019). It is important to consider design education's role in perpetuating and reifying inequity (Mazzarotto & Serpa, 2022). Too often design research and solutions focus on users who are WEIRD, i.e., Western, Educated, Industrialized, Rich, and Democratic (Coney, 2022; Mustafa, 2023), and scholars have also critiqued exploitative practices of approaching marginalized communities for class projects without appropriate reciprocity, and there has been an increased emphasis on decolonial design and participatory co-design (Bennett & Rosner, 2019; Ymous et al., 2020).

In addition, it is important to examine the diversity of the design field, which starts with examining education and who we are training to be 'designers'. The stated pedagogical purpose of design education is to help students "understand what it means to be a designer," (Shaffer, 2007) and while departments all over the world are minting thousands of designers annually through such courses, a closer examination shows how the model of design education we practice today is deeply ableist and exclusionary, most often towards students with traditionally marginalized identities.

Before talking about these ableist and exclusionary practices in design education, it is important to briefly address the language around disability and disabled people. There are a variety of preferences in the disability community for either person-first language, i.e. "people with disabilities", and identity-first language, i.e. "a disabled person" (Dunn & Andrews, 2015). Out of respect to the diverse preference of the disability community and recognizing that there is no singularly preferred language across different sub-communities, we will use both of these versions interchangeably.

Lesley-Ann Noel (2021) defines exclusion as the act of "deny[ing] a person or group access to a place, a group, or a privilege," either intentionally or unintentionally. While there is a lot of focus on intentional exclusion and rightfully recognizing the malevolence they are often born out of, unintentional or absent-minded actions that become exclusionary are harder to think about. Such unintentional exclusion almost happens instinctively, likely because of tendencies ingrained within individual cultures and histories (Noel & Paiva, 2021), which lead to the practice of ableism. Talia Lewis (2022) defines ableism as "a system of assigning value to people's bodies and minds based on societally constructed ideas of normalcy, productivity, desirability, intelligence, excellence, and fitness... This systemic oppression that leads to people and society determining people's value based on their culture, age, language, appearance, religion, birth or living place, health/wellness, and/or their ability to satisfactorily re/produce, 'excel' and 'behave.' You do not have to be disabled to experience ableism." Towards developing an inclusive and non-ableist practice, it is important to first grow a habit of learning to recognize exclusive and ableist actions when they occur. Let us explore some of these actions in design studio classrooms.

Design education operates through a constructivist model of learning (Papert, 1980), and the principle that students learn best by *doing* design. In our classrooms, design instruction is thus most often delivered in the studio model, which operates through collaborative group or parallel individual work sessions (Ata & Dogan, 2021). These studios are often multiple hours: e.g. in our department most graduate studios are 6-10 PM. However, some of the key assumptions behind this structure are ableist. As a start, especially before the advent of the COVID-19 pandemic, such sessions would mostly be face-to-face, which is exclusionary to the non-typical college student, or anyone with chronic illness, juggling multiple jobs, parenting, etc. A fully face-to-face design studio course would force them to find workarounds to their own unique situations to be present in class, often at cognitive, physical, emotional and/or financial cost, disadvantaging their participation before the class even begins.

Furthermore, the mechanics of face-to-face participation and group work also present in ways that are exclusionary. Group design sessions typically involve a lot of visual and auditory communication, which makes it difficult for D/deaf or hard-of-hearing (DHH) and/or blind and low-vision (BLV) students to fully participate in design conversations. This can manifest in many different ways. For DHH students, difficulty to participate in group conversations can be amplified by the chatter from conversations in other groups in close proximity within classrooms. If they are working with a sign-language interpreter or captioner to participate in group conversations, focusing there will detract from their abilities to follow along with visual activities such as affinity mapping or codesigning (Ang et al., 2022; McDonnell et al., 2023). Such design sessions with strong emphasis on visual participation are also inaccessible for BLV students, especially when it comes to activities such as building physical prototypes by hand (Stangl & Yeh, 2019). BLV students might also struggle with co-analysing user interviews, which usually happen through collaborative examination of interview transcripts or notes, since the generated transcripts might not be screenreader friendly or notes taken by other interviewers will not be accessible to them. Students with red-green colorblindness might also struggle in affinity mapping or other exercises involving post-its or sticky notes, since red and green are two of the most common colors they come in and activities like affinity mapping often use colors of stationery to encode some meaning.

The normative pedagogical practices in design studio courses can also be exclusive to students with a wide range of cognitive and intellectual disabilities. Neurodivergent students (e.g. ADHD, autism, sensory processing disorder) might not be able to fully focus on small group conversations in face-to-face design studio courses, being distracted or overstimulated by other group conversations in the classroom, among several other possible outcomes which will be unique to each student. Students with aphasia might have trouble contributing to such conversations, or conducting the user research or user testing interviews often mandatory in design courses. Students with anxiety or depression might not be fully present in class conversations during groupwork, and therefore reduce the collective productivity of the group to incur displeasure from their group members or teaching staff. Over and above this definitely non-exhaustive list, students with multiple marginalities, e.g. a student who is both DHH and has ADHD, will be marginalized and excluded from studio sessions in many more ways than we can possibly list in this section.

All of these operate under the general institutional climate, although slightly improved in the past decade, that disability accommodations in classrooms are unnecessary beyond a certain point, and ultimately a way for conniving students or their parents/guardians to acquire extra academic advantages ahead of their peers (Worth, 1999). The "certain point" here usually lands on the line between visible and invisible disabilities: while universities and professors are willing and able to accommodate the needs of a wheelchair user or someone on crutches with special seating or other approaches, they are less likely and able to accommodate students with ADHD with potential coping mechanisms like fidgeting during class or leaving an earbud in with light music playing, viewing such behavior as deviant from classroom norms (Kofler et al., 2008; Orban et al., 2018). While this is generally true for college education as a whole, it applies more specifically in design studio courses, where the normative accommodations for students with common invisible disabilities might not be necessary due to course structures, thus lulling instructors into thinking their classrooms are already accessible and they need not do any further work. For instance, one of the most common accommodations for students with ADHD or dyslexia is extra time and a separate location for midterms and final examinations, which are fairly uncommon in design studio courses. Therefore, an instructor receiving a notification from Disability Services that says the student needs exam accommodations might not think beyond the fact that their course does not involve exams, and therefore miss opportunities to intentionally make their courses more accessible for the student.

Thus, the current pedagogy for design education, even while teaching principles of diversity, equity and inclusion in classes, is rife with practices that are ableist and exclusionary, often towards traditionally and multiply marginalized students. In the following sections, we explicate three case studies of courses we were assigned to teach, where we adapted course structures and materials to accommodate students who would otherwise have been excluded under existing course design. We discuss our practices for accommodation, the successes and failures, and lessons learned for future course design.

## 3. Methods

We base the construction of the upcoming three case studies based on our own experiences as instructors of these courses. We adopt an autoethnographic methodology similar to our previous work (2022, 2023) and that of others (Glazko et al., 2023; Mack et al., 2023), in reflecting upon our experiences as instructors after the culmination of each course. We rely on individual documentations and notes from course experiences, rather than formally collected data from students or other sources.

While we believe that the experiences and practices we document below are ubiquitous across design studio courses beyond just the ones at our University, we acknowledge that the findings presented might have room for disagreement from other instructors of similar courses with positionalities different than our own. The first author is a current doctoral candidate as an international student to the US, self-identifying as able-bodied, neurodivergent, and a man of color, with 4 years experience teaching design studio courses. The second author is a teaching faculty member at an R1 university, self-identifying as a disabled white woman, with 10 years experience teaching such courses.

## 4. Case Study 1

The first case study discussed here is an undergraduate-level Introduction to User-Centered Design (BUCD) course, taught by one or both of the co-authors for seven consecutive quarters. Typically, this course consists of 30-40 university sophomores and juniors who have just been admitted into the major, and take this as one of their first major courses. In a ten-week quarter system, students go through the entire user-centered design process (Norman, 2013), from idea formation and user research through to multiple rounds of designing and user-testing to deliver a high-fidelity prototype. Like most other design courses, it is typically taught in a flipped classroom format with studio components (Jared et al., 2014; Koutsabasis & Vosinakis, 2012), with face-to-face attendance and class participation being integral to course success, at least prior to the Covid-19 pandemic. Informed by the principles of Universal Design for Learning, there are no timed tests, and there are grace periods for deadlines.

One of the first quarters the authors co-taught this class was when universities across the US were returning to face-to-face learning. In such a climate, we were encouraged by university policies to teach the class fully face-to-face with some asynchronous online participation opportunities and options in case students fall sick. However, instead of building an face-to-face course structure and setting up remote attendance as an afterthought akin to most systems designed for default users and retrofitting affordances for accommodations (Dolmage, 2017), we instead decided to adopt a HyFlex teaching structure (Beatty, 2014) with synchronous face-to-face and remote participation opportunities, and allowed students to make a day-to-day choice on their mode of attendance without requiring instructor approval or any prior communication. Such an approach is known to be successful in achieving course objectives in design studios and other types of courses (Ghosh & Coppola, 2022, 2023; Kim, 2023; Mentzer et al., 2023; Miller & Baham, 2018), while also making courses more accessible to accommodate a wide range of individual student needs.

However, a HyFlex course structure isn't trivial to adopt, and is much more involved than simply opening up a Zoom meeting in a face-to-face class. Especially in a studio based class, it is about trying to simulate the face-to-face experience as closely as possible, and not just ensuring that students can hear and follow along with lectures remotely. We attempted to do so by utilizing hybrid meeting technology in the form of a Meeting Owl 360, an audio-video device that captures conversations within a 10ft radius and displays multiple fields of view. We also adopted course materials, which were traditionally intended to be used in either fully face-to-face or fully remote formats, to a HyFlex structure, such that students would have equal modes of participation irrespective of their mode of attendance. This meant switching brainstorming and ideation sessions away from physical sticky notes to Miro boards, preparing interactive handouts that could be used in both modalities, and coming up with remote-friendly versions of design and sketching activities where students in the classroom can leverage provided stationery and materials. Furthermore, the group work component of the class also needed to be addressed on a daily basis because student attendance format could change on any given day. We did so by paying attention in every class period to which students were face-to-face or remote in every group, and individually ensured that every group member was looped into conversations, by asking face-to-face students to use their own devices to join remote teammates on zoom.

While intended as an accommodation for temporary illness, HyFlex afforded flexibility and access for a variety of situations. Two of the students had serious injuries the first week of class that would make face-to-face attendance impossible for them. They were able to participate from home, as opposed to having to drop the course as they were advised to do in other courses that were either fully face-to-face. Students maintained an almost-perfect attendance record, did well in the class, and delivered a high-quality prototype as part of their final assignment, later communicating their appreciation for the HyFlex format as an effective form of class participation and achieving their own course goals (redacted for review). Indeed, the HyFlex modality also provided access for one of the authors when they sprained their ankle and were unable to commute to campus.

The HyFlex model was also successful in accommodating a variety of student circumstances across the seven quarters the authors were associated with the course, chronic illness that made 9:30 AM courses difficult, mental health issues, personal bereavement, and unforeseen travel circumstances. Therefore, it can be said with unequivocal confidence that this model of instruction made our BUCD class spaces more accessible, and its absence would have made the spaces exclusionary for both the aforementioned students and the instructors themselves.

## 5. Case Study 2

The second case study we present is a Masters-level Introduction to User-Centered Design (MUCD) course, taught by the second author. Similar to the BUCD course described above, it features a quarter-long group design project and is taught in the HyFlex format. This class is taken by all incoming Masters students who are split into multiple 30-40 person studios. The course emphasizes visual design with weekly sketching assignments and group assignments to make visual storyboards, journey maps, and user interfaces.

This class presents an access mismatch for blind and low vision (BLV) students. We were surprised to learn that few resources existed for teaching BLV designers, and there were no software packages that afforded independent design work for BLV designers (Shinohara et al., 2022). Unfortunately, we received little support from the institutional disability services. This creates a systemic barrier that prevents BLV students from design professions and propagates ableist and inaccessible designs.

To make the class more accessible, we brought in experts on tactile graphics as guest speakers, adapted some of the sketching assignments to be tactile or auditory, required auditory podcast style storyboards, and challenged every team in the class to create a final prototype that was not screen-based. We chose assigned readings that were available in digital formats, though most did not have image alternative text. We required all final presentations to have captions and alt text, and we provided tactile diagrams when relevant.

The adaptations we made to this course did not fix every mismatch and were not universally successful, but they were an important first step towards including all of our students (publication forthcoming). This experience also provided an important educational opportunity for non-disabled students to forefront accessibility in design, and the adaptations will remain in the course regardless of if there are BLV students enrolled.

## 6. Case Study 3

In the third case study, we discuss a Masters-level Accessibility and Inclusive Design (AID) course, co-taught by the authors in one quarter and individually by the second author in two other quarters. This course addresses different forms of human diversity and how existing systems marginalize those with certain attributes of each identity, with a consideration of how students, as budding designers, can do better and adopt more inclusive practices. The studio component asks them to co-design with a user/expert participant who has a lived experience of disability (IHCD, 2022; Niedderer et al., 2022).

As part of a professional Master's program, classes were four hours in the evening to accommodate the fact that many students were full-time industry professionals. However, we recognized right off the bat that attending face-to-face classes at this time in the winter weather might not be accessible to everyone. Such a mode of instruction might be inaccessible to parents, especially single-parents, who typically can only access daycare during working hours, or those who have long commutes via public transit. Furthermore, the night hours and late end time also makes it difficult for wheelchair users or people with ambulatory aids to navigate campus pathways that are dimly lit and limit food options after the closure of on-campus cafes and restaurants, thus making the requirement of in-person attendance an exclusionary practice (Hamraie, 2023). Therefore, we adopted a similar HyFlex structure as above, to offer students the ability to attend class remotely and still have the full course experience without requiring to be on campus until late into the evening.

To model inclusive practices in a class on inclusivity, we began by co-designing some class norms and building accessible practices such as using auto captions on personal devices around the room and always stating names when speaking. Educational disability design projects can be extractive (Jackson et al., 2022), so users/experts were paid for their participation, and students were trained on ethical research and appropriate methods for working with disabled people (Mack et al., 2022; Protections (OHRP), 1979; Spiel et al., 2018; Williams & Gilbert, 2019). Students were reminded to respect the expertise of the user/expert and to share power in design decision making. In addition, there can be tension when students feel they need something for a grade, so we used a form of ungrading for the course, i.e. where we have conversations with students about the grade they believe they have earned in the class based on their achievement of course objectives (Blum, 2020; Coppola & Turns, 2023).

Despite our efforts, we were structurally limited by the design of our classroom and the technology provided. The room's acoustics made it difficult for the live captioner and people on zoom to hear speakers. This was particularly problematic when we had DHH guest speakers who relied on captioning or those with speech disabilities. The campus, building, and room's inaccessibility made it difficult to have guest speakers with mobility disabilities, and the furniture in the room further exacerbated the access conflicts when DHH students could not hear disabled guests. For group work, we were able to mitigate the sound issues by sending teams out into unused classrooms where they could hear better and/or use captioning more reliably. Unfortunately, we tried to host a poster showcase on the last day of class, and our many DHH students, users/experts, and guests were unable to participate because it was too loud to use captioning in the room.

## 7. Towards a Pedagogy of Care in Design Education

Based on our experiences in teaching and adapting the aforementioned courses through a combination of different strategies to address the multiple ways in which existing course setups were ableist and exclusionary to our students, to varying degrees of success and failure, we build towards a pedagogy of care in design education to center the needs of the traditionally marginalized identities in design classrooms. Our proposals embody the care ethics principles such as Olena Hankivsky's (2005) "commitment to provide the opportunity and a safe space for others to express their 'otherness'," Tony Monchinski's (2010) urge to "realistically differentiate instruction by tailoring pedagogy to the specific needs of individual students," and Hamington and Flower's (2021) directive to frame practice as "a response to the particularity of someone's circumstance that requires concrete knowledge of their situation, entailing imaginative connection and actions on behalf of their flourishing and growth," as well as more foundational ideas of scholars like John Dewey (1938), Carol Gilligan (2016), Nel Noddings (1984) and Joan Tronto (2005), which frame pedagogies of care around everyday principles of recognizing one's own positions of power and precarity alike, and consistently supporting those with less power with meaningful and intentional actions.

Disability scholars have critiqued pedagogical norms and have proposed pedagogies of care in fields such as English Composition (Burtis & Quinn, 2022). Currie and Hubrig (2022) propose that structural course design choices such as flexibility, co-created syllabi, and forefronted accessibility can shift the burden off disabled students and disrupt ableist pedagogies. We echo these, and further advocate that design education needs a pedagogy of care that centers access and inclusion for all students as an important first step towards educating a more diverse design workforce and a more equitable and inclusive world.

### 7.1 Adopting a HyFlex Model of Studio Courses

The first component of our pedagogy of care in design education is to practice HyFlex models of design studios. While some researchers argue that the absence of an face-to-face component negatively impacts student grades and learning (Green, 2021), we align our own experiences teaching HyFlex courses over three years with prior research (e.g. Ghosh & Coppola, 2022, 2023; Mentzer et al., 2023; Miller & Baham, 2018; Sowell et al., 2019 and others) to state that not only do HyFlex courses deliver on course expectations and student metrics, but also they accommodate the many diverse forms on a case-by-case basis. Therefore, towards developing a pedagogy of care in design education that resists the existing ableist and exclusionary trends in the field, we believe that HyFlex versions of design studio courses go a long way in the right direction (van Kampen et al., 2022).

We also reaffirm that a HyFlex model is not just opening up a synchronous Zoom room for students to log in, but instead must be executed with a care to adapt course materials for equal synchronous remote and face-to-face participation and provide students with the day-to-day control to independently determine their chosen mode of course participation, without seeking permission, to give them the agency to practice self-care (Kabeer, 1999).

### 7.2 Offering Non-Visual Participation Options

As mentioned before, design studio courses rely heavily on visual elements of design and sketching, which presents an access challenge for BLV students. We tried a few adaptations in the MUCD course, to some degrees of success, but there is more work to be done.

For a discipline so centered around sketching the everyday objects sighted designers can see in their daily lives, there have to be more options to accommodate their BLV counterparts. We recognize the work of Siu et al. (2021), whose tool of sketching augmented by audio and haptic guidance is an important contribution in this space. There are also avenues to draw inspiration from the fields of arts education, and their efforts to provide meaningful access to the inherently visual aesthetics for BLV artists (De Coster & Loots, 2004). We also advocate for co-designing such options with BLV students, valuing them as experts of their own lived experiences (Hohl et al., 2022; Williams & Gilbert, 2019).

We are continually committed to exploring adaptations to existing course structures for BLV students, and are currently in the process of employing some in a different course being taught by the second author which has BLV students.

### *7.3 Seeking Institutional Support*

Though we discussed our own efforts in reworking existing ableist and exclusionary course structures and radically designing inclusive practices that accommodate a wide range of student needs, we would be remiss to ignore the overwhelming amounts of effort it took to do so, both each time and cumulatively across many classes. In cases such as painstakingly setting up and taking down the technology for HyFlex classrooms, and reworking entire course loads of materials and assignments, the work required to make our courses more inclusive was often conducted on our own times, and very much outside the purview of job expectations. We did so because we were committed to designing more inclusive spaces, but it would be grossly unfair of us to expect such degrees of work from everyone assigned to teach such courses. Indeed, junior instructors such as us committed to redesigning existing ableist structures into more inclusive ones often find themselves in situations where they want to make good changes but lack the required time and resources to do so mindfully.

There is also a doubly-precarious situation of power dynamics, where instructors such as ourselves are too low-power to individually demand university resources under the misguided narratives that accommodations are misused by opportunistic students or that they water courses down to hold back the academically gifted (Dolmage, 2017), but also are under pressure both from students and our own consciences to practice care ethics of teaching through empathy and action that is "a response to the particularity of someone's circumstance that requires concrete knowledge of their situation, entailing imaginative connection and actions on behalf of their flourishing and growth." (Hamington & Flower, 2021). Therefore, a sustainable pedagogy of care in design education cannot simply focus on students, but also demonstrate empathy and support towards the instructors and teachers who hold up university educational systems (Bennett & Rosner, 2019; Piepzna-Samarasinha, 2018). Ableism is a system of oppression, and it can only be solved by systemic solutions.

## 8. Limitations and Future Work

One potential limitation of this work is that we present this work entirely from our perspectives as instructors of design courses within the field, and do not conduct interviews with current and past students for their opinions on this matter. For future work, we are interested in conducting a wider-scale exploration of this topic involving students across the Bachelor's and Master's levels, as well as with other instructors. As of this writing, such work is ongoing and under submission. Further future research could also extend this across different universities and educational contexts, and students with other disabilities than the ones we focused on.

## 9. Conclusion

In this paper, we argue that the prevailing pedagogies of design education are ableist and exclusionary towards students with disabilities and cast doubt upon the question of who such classes train to be called "a designer". We have presented three case studies in which we navigated through accessibility issues in our courses, and the pedagogical decisions we made in order to improve access. These included allowing remote participation, adapting materials to be more accessible for BLV, DHH, and neurodivergent learners, flexible deadlines, ungrading, and co-created class norms. These choices were done within our own context with high instructor-student ratios and access to videoconferencing technologies.

While we improved accessibility and inclusion in some ways in the aforementioned courses, we do not pretend that we have perfectly fixed every access mismatch or solved the structural ableism inherent in the studio model of education. Readers of this work and others in the field should consider this a small step towards more equitable and inclusive practices in design education, and strive towards making their own additions and designing adaptations in their own courses driven by student needs. Abolition of existing ableist practices is a collectivist, community-driven approach, and together, we can work towards building course structures that are inclusive of all backgrounds, contexts, and identities.

(2020). "I am just terrified of my future"—Epistemic Violence in Disability Related Technology Research. *Extended Abstracts of the 2020 CHI Conference on Human Factors in Computing Systems*, 1–16. https://doi.org/10.1145/3334480.3381828

About the Authors:

**Sourojit Ghosh** is a 4th year PhD Candidate in Human Centered Design and Engineering at the University of Washington, Seattle. His research interests are around equity in engineering, as he studies how large systems driven by machine learning algorithms such as social recommender systems and generative AI tools perpetuate societal harms around traditionally marginalized peoples. Parallely, he studies how engineering pedagogy can become more equitable, particularly to include students with disabilities and other identities marginalized in such spaces.

**Sarah Coppola** is an Assistant Teaching Professor at the Department of Human Centered Design and Engineering at the University of Washington, Seattle. An educator and researcher, her work focuses on how technology and systems design affects people's performance and health. She holds a BS in Mechanical Engineering from Northwestern University, an MS in Human Factors Engineering from Tufts University, and a Doctorate in Ergonomics from Harvard University.